\documentclass[aps,rmp,reprint,amsmath,amssymb,graphicx,longbibliography,floatfix]{revtex4-1}
\usepackage{amssymb,amsfonts,amsmath} 
\usepackage{graphicx}
\usepackage{color}
\usepackage{xcolor}
\usepackage{graphics}
\usepackage{bm}
\usepackage{booktabs}
\usepackage[breaklinks=true]{hyperref}
\hypersetup{bookmarksnumbered, pdfpagemode=UseOutlines, 
pdfdisplaydoctitle, 
colorlinks=true, citecolor=blue, filecolor=blue, linkcolor=blue, urlcolor=blue}

\newcommand{\moire}{moir{\'e}} 

\begin{document}

\title{Review: Advanced characterization of the spatial variation of \moire{} heterostructures and \moire{} excitons.}

\author{Alberto de la Torre}
\affiliation{Department of Physics, Northeastern University, Boston, Massachusetts, 02115, USA}
\affiliation{Quantum Materials and Sensing Institute, Northeastern University, Burlington, Massachusetts, 01803, USA}

\author{Dante M. Kennes}
\affiliation{Institute for Theory of Statistical Physics, RWTH Aachen University,and JARA Fundamentals of Future Information Technology, 52062 Aachen, Germany}
\affiliation{Max Planck Institute for the Structure and Dynamics of Matter, Center for Free Electron Laser Science, 22761 Hamburg, Germany}
\author{Ermin Malic}
\affiliation{Fachbereich Physik, Philipps-Universität Marburg, 35032 Marburg, Germany}
\affiliation{Department of Physics, Chalmers University of Technology, Gothenburg, Sweden}
\author{Swastik Kar}
\affiliation{Department of Physics, Northeastern University, Boston, Massachusetts, 02115, USA}
\affiliation{Quantum Materials and Sensing Institute, Northeastern University, Burlington, Massachusetts, 01803, USA}
\affiliation{Department of Chemical Engineering, Northeastern University, Boston, Massachusetts, 02115, USA}

\date{\today{}}

\begin{abstract}
In this short review, we provide an overview of recent progress in deploying advanced characterization techniques to understand the effects of spatial variation and inhomogeneities in \moire{} heterostructures over multiple length scales. Particular emphasis is placed on correlating the impact of twist angle misalignment, nano-scale disorder, and atomic relaxation on the \moire{} potential and its collective excitations, particularly \moire{} excitons. Finally, we discuss future technological applications leveraging \moire{} excitons.

\end{abstract}

\maketitle
\tableofcontents

\section{Introduction} %General Intro

\begin{figure*}[t!]
\includegraphics[width=\textwidth]{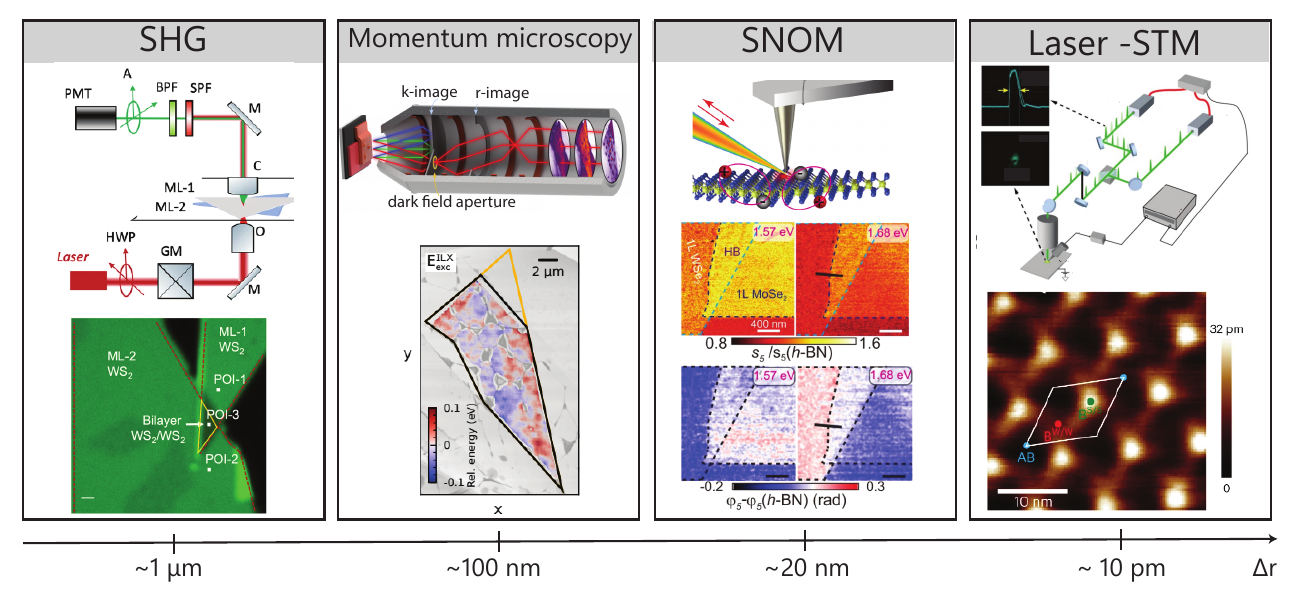}
\caption{A subset of advanced characterization techniques deployed to study the spatial variation of \moire{} heterostructures. From left to right, in order of increasing resolution, Second Harmonic Generation (SHG) ($\sim 1 \mu$m), photoelectron momentum microscopy ($\sim 100$~nm), scanning near-infrared optical microscopy (SNOM,  $\sim 20$~nm), and laser scanning tunneling microscopy and spectroscopy (laser-STM/S, $\sim 10$~pm). Adapted with permission from \cite{Psilodimitrakopoulos2021,Zhang2022,Mogi_2019,Li2024,schmitt2023ultrafast}
\label{fig1_Intro}}
\end{figure*}

The emergence of two-dimensional van der Waals (v.d.W.) materials has raised the intriguing idea of mixing and matching layers of different materials to realize novel properties in a given stack \cite{geim2013van}. Moir{\'e} patterns emerge when two or more atomically thin v.d.W. layers are artificially stacked with a slight angular misalignment or, in the case of heterolayers, featuring a small lattice constant mismatch \cite{Du_Science_2023}. Remarkably, the \moire{} interference patterns create a superstructure, the \moire{} lattice, with a new periodicity usually much larger than the underlying atomic lattices. This superlattice can dramatically alter the electronic properties of the heterostructure. The long spatial periodicity of the \moire{} potential leads to the formation of minibands that directly modify the transport behavior. Some key discoveries in this field of \moire{} engineering include the formation of flat bands \cite{MacDonald_flat_Bands}, the enhancement of correlations \cite{Zhang2022_Feng}, the engineering of topological properties \cite{Choi2021}, the formation and control of polaritons \cite{Zhang2021_polaritons} or networks of solitons \cite{Fu2020}. Moir{\'e} heterobilayers also enable the creation of excitons \cite{Jin2019NAt}, the key quasi-particle we will focus upon in our review. We refer the reader to recent reviews for an in-depth discussion of \moire{} physics: \cite{Behura2021,Alexeev2019,Mueller_excitond_device_2DTMD,Huang_Excitons_moire_superlattices,Ciarrocchi_Excitonic_devices_vdWheterostructures,Controlling_QM,Liu2021}. The mechanical properties of the v.d.W. compounds in the heterostructure are also profoundly impacted due to the \moire{} potential. While the effect of the \moire{} superlattice on the tribological properties of the interface is not the focus of this review, we refer the readers to the in-depth discussion presented in \cite{Yan2024}.

Moir{\'e} excitons refer to photo-excited electron-hole bound states that form within (intra-) or between (inter-) layers with the potential defined by the \moire{} superpattern \cite{Tran2019}. Moir{\'e} excitons exhibit distinct energy levels and behaviors compared to their semiconductor-monolayer or three-dimensional counterparts. Moir{\'e} excitons can host a rich tapestry of phenomena with valuable insights into novel quantum phenomena and many-body effects in low-dimensional systems, including exciton condensation, superfluidity, and topologically protected states \cite{Shimazaki2020, Choi2021,Qi2023}. Moir{\'e} excitons also offer opportunities for tuning the electronic band structure of materials, enabling tailored properties for semiconductors, insulators, and topological materials, paving the way for technological innovation \cite{Ciarrocchi_Excitonic_devices_vdWheterostructures,Qi_review_2022}. Additionally, they contribute a new dimension to valleytronics, allowing for control over valley polarization and dynamics in v.d.W. heterostructures \cite{Unuchek2019,Jin2019}. Furthermore, \moire{} excitons play a pivotal role in realizing topological phases and the quantum anomalous Hall effect, leading to the development of innovative electronic and spintronic devices \cite{Li2021_QAH}. Interactions between \moire{} excitons and antiferromagnetic ordering enable the manipulation of spin-dependent properties and spin information in optoelectronic devices \cite{Ramos_2022_PL_enhancement}. Moreover, \moire{} excitons present opportunities for controlling carrier dynamics and charge transport properties in heterostructures, resulting in high-performance optoelectronic devices with improved efficiency and reduced energy consumption. Excellent, complementary reviews of \moire{} excitons are \cite{Huang_Excitons_moire_superlattices,Regan2022} focusing on early optical experiments and selection rules. Moreover, \cite{Ciarrocchi_Excitonic_devices_vdWheterostructures} and \cite{Mueller_excitond_device_2DTMD} highlight device architectures and engineering approaches to tailor excitons toward future applications.

Controlling and understanding \moire{} superlattices, arising from the periodicity of \moire{} patterns, is fundamental to engineer \moire{} excitons and their dynamics \cite{brem_2020,Merkl2020,meneghini2024excitonic}. However, the excitonic properties of \moire{} heterostructures are extremely sensitive to local deviations from perfect twist angles and Lego-like stacking description. In this short overview, we discuss how new theoretical descriptions and advanced characterization techniques have been deployed to study the role of spatial variations over multiple length scales, namely twist angle misalignment, local inhomogeneities, and atomic relaxation, affect the \moire{} potential and the physics of \moire{} excitons. We conclude by discussing possible technological applications based on integrable \moire{} devices that motivate further efforts to understand and control local spatial variations.

\section{Characterizing the effects of spatial variation on the \moire{} potential and \moire{} excitons by advanced tools.}

Excitonic states in \moire{} v.d.W. superlattices offer exciting prospects for novel optoelectronics and quantum technologies \cite{Ciarrocchi_Excitonic_devices_vdWheterostructures}.  The strong exciton physics in transition metal dichalcogenides (TMDs) has been extensively investigated using different optical spectroscopy techniques, such as absorption spectroscopy, photoluminescence and photocurrent measurements, and optical pump-probe measurements \cite{Huang_Excitons_moire_superlattices}. While these techniques provide access to spectral and temporal degrees of freedom, they lack information about the real space properties of excitonic TMDs. However, local spatial modulations of the \moire{} superstructure due to strain fields, atomic defects \cite{pnas_interplay_defffect_moire}, and twist angle change extremely influence their optoelectronic properties. Correlating the averaged optical and transport properties with these local variations is needed to unambiguously resolve the role of the \moire{} energy and length scale. 

In the following, we review recent applications of advanced characterization techniques to disentangling emergent excitonic phenomena from \moire{} potential disorder in order of spatial resolution. In particular, as highlighted in Figure \ref{fig1_Intro}, we focus on Second Harmonic Generation (SHG), momentum microscopy, atomic force microscopy (AFM) based techniques, and laser Scanning Tunneling Microscopy (STM). In Section \ref{subsection:optics}, we discuss how the extreme sensitivity of SHG, a non-linear optical technique, to inversion symmetry breaking enables quick and non-intrusive detection of deviations from the twist angle in \moire{} heterostructures \cite{Psilodimitrakopoulos2021}. We also review how scattering Scanning Near-field Optical Microscopy (s-SNOM), a technique that combines the spatial resolution of AFM with optical spectroscopy, has been deployed to study variations in the excitonic response with nanometer resolution \cite{Zhang2022}. In Section \ref{subsection:ARPES}, we review the role of time-resolved Angle Resolved Photoemission Spectroscopy (tr-ARPES) in characterizing the formation dynamics of hybrid \moire{} excitons \cite{Menghini_2023}. We also discuss how ultrafast dark field momentum microscopy, a recently introduced technique, reveals the role of local spatial inhomogeneities in the charge transfer dynamics of \moire{} excitons \cite{schmitt2023ultrafast,Schmitt2022}. Finally, in Subsection \ref{subsection:STM}, we discuss how piezoresponse force microscopy (PFM), an extension of AFM, and STM have been deployed to study the \moire{} potential and deviations from the ideal membrane-like behavior of the layers \cite{Halbertal2022}. Moreover, the extension of STM to the time domain by its combination with laser pulses \cite{Mogi2022} has revealed the role of corrugation and strain in the lifetime of \moire{} excitons. 

\begin{figure*}[t!]
\includegraphics[width=\textwidth]{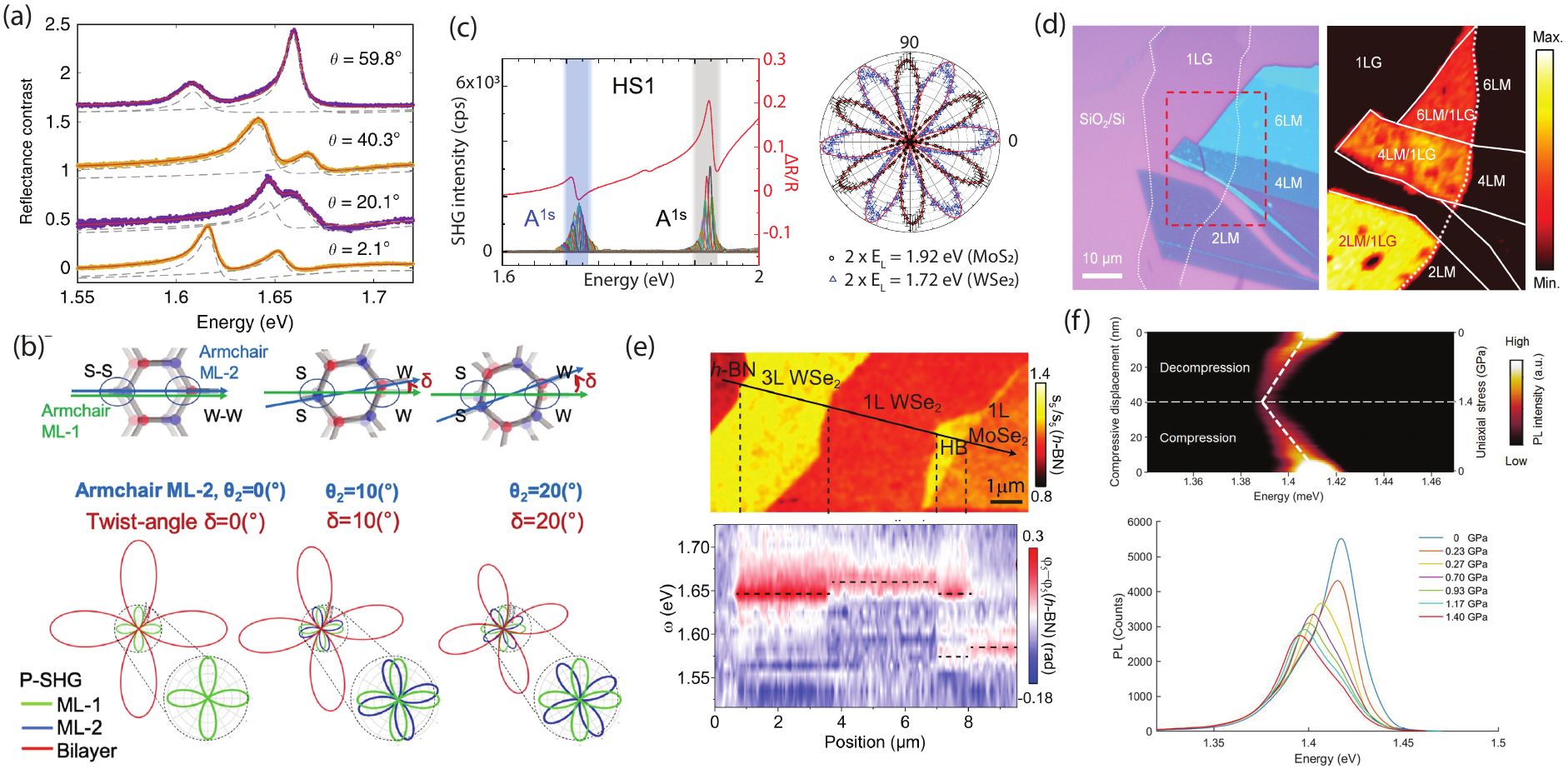}
\caption{ (a) Normalized reflection spectroscopy of WS$_2$/MoSe$_2$ bilayers as a function of twist angle. (b) Schematic of the alignment of the main crystallographic directions as a function of twist angle, $\theta$, keeping the armchair direction of the bottom monolayer (ML) fixed and rotating the armchair direction of the top ML. The simulated interference SHG pattern shown in the top row reflects the twist angle by the rotation of the lobes and the modulation of the overall intensity. (c) SHG spectroscopy on a WSe$_2$/MoS$_2$ heterostructure with $\theta = 0^{\circ}$. The polar plot shows the SHG pattern at the energy of the WSe$_2$ exciton (1.72 eV) and that of MoS$_2$ (1.92 eV). (d) Micrograph of MoSe$_2$ (LM) and Graphene (LG) for different stacking configurations and the associated second-harmonic generation (SHG) spatial imaging map in the highlighted dashed red box. (e) Normalized phase spectra at different positions in the sample indicate the excitonic responses for the WSe$_2$, MoSe$_2$ monolayers, and the MoSe$_2$/WSe$_2$ heterobilayer. (Top) PL spectra of the WSe$_2$/WS$_2$ interlayer exciton evolution under applied stress. (Bottom) The horizontal line cuts across the top panel showcase the shift towards low energies of the interlayer exciton. (f) Panels adapted with permission from \cite{Zhang2020_twist,Psilodimitrakopoulos2021,PhysRevB.105.115420, SNOM_pressure,Zhang2022}.}
\label{fig1_SHG}
\end{figure*}

\subsection{Non-linear optical techniques deployed to \moire{} heterostructures.}\label{subsection:optics}

In the fabrication of heterobilayer devices, the twist angle, $\theta$, the effective misalignment between the main crystallographic directions of the two stacked monolayers, has emerged as a novel degree of freedom that controls the \moire{} potential and enables tunability of the optical emission and absorption spectrum of TMD heterobilayers \cite{Huang_Excitons_moire_superlattices}. In general, perfect stacking can be achieved at $\theta = 0^{\circ}$ (AA stacking) and $\theta = 60^{\circ}$ (AB stacking). The almost perfect overlap between the Brillouin zones of each monolayer can lead \moire{} inter- and intra-layer excitons. The onset of excitonic correlations depends on electrostatic doping, on the v.d.W. materials participating in the heterostructure, and on twist angle \cite{Tang2020, Barr2022}. For example, perfectly stacked heterostructures of WSe$_2$/WS$_2$ display more strongly interacting hybrid excitons than those of MoSe$_2$/WS$_2$ due to differences in the local TMD dielectric constants \cite{Tang2020}. Small deviations from the AA or AB stacking lead to large changes in the band overlap in momentum space and a reconfiguration of the excitonic energies. This is illustrated in Figure \ref{fig1_SHG} (a), where reflection spectroscopy measurements of the lower ($\approx 1.6$~eV) and upper hybrid excitons ($\approx 1.65$~eV) revealed changes to the excitonic spectrum of more than $100$~meV as a function of the twist angle \cite{Zhang2020_twist}. Optical absorption spectroscopy has shown a similar dependence of the excitonic spectrum on the twist angle in heterobilayers of WSe$_2$/MoSe$_2$ \cite{Barr2022}. Thus, precisely determining the stacking angle between the heterobilayers is fundamental for future devices based on \moire{} excitons. In this context, the observation of strong non-linear responses dependent on the twist angle has emerged as a unique and versatile tool to characterize the formation of the \moire{} excitons \cite{Jin2019,Tran2019}. Similarly, changes of symmetry and nonlinear responses associated with the interlayer coupling and charge transfer as a result of the \moire{} potential have also been detected via non-linear responses \cite{Shree2021_Guide_NatPhysRev,PhysRevB.106.155410,YANG20201361}.

For strong electric fields, the interaction of light can be modeled by a resulting polarization given by:
\begin{equation}
 \begin{split}
       P_i(t) = \varepsilon_0 (\chi^{(1)}_i E_i(t) + \chi^{(2)}_{ij}E_i(t)E_j(t) +\\
       + \chi^{(3)}_{ijk} E_i(t)E_j(t) E_k(t) +...)
 \end{split}
\end{equation}
with $\chi^{(i)}$ as the linear and non-linear optical susceptibility tensors and $\varepsilon_0$ as the permittivity of vacuum. These high-order processes are governed by the crystallographic symmetry and the electronic band structure, carrying fundamental information about the wavefunctions involved in the two-photon absorption process \cite{Ma2021}. Terms with an even number of powers on the electric field $E$ are only allowed in inversion symmetry-breaking compounds. The second-order term carries special importance as it is responsible for second-harmonic generation (SHG), in which the constructive interference of the two oscillating electric fields with frequency $\omega$ can give rise to a polarization density at twice the frequency $P(2\omega)$ \cite{zhao2018second}. In a normal-incidence SHG experiment, a microscopic objective focuses a single wavelength or broadband light pulse with controlled polarization onto the sample. Emission from the sample is first collimated by the objective and short-pass filtered to eliminate contributions from the fundamental beam before being focused onto a photodiode or CCD camera. Resolving the underlying structural point group can then be achieved by tracking the angular dependence of the intensity of the $2\omega$ response with respect to the main crystallographic symmetries and the polarization of the incoming and outgoing beam.

For 2D TMD monolayers, a strong SHG response with maxima along the armchair or zig-zag direction can be measured due to their inversion symmetry-breaking nature (point group $D_{3h}$). When a heterobilayer is formed, the SHG signal from each layer coherently interferes, resulting in a pattern that is extremely sensitive to small twist angle misalignment from perfect AA and AB stacking, as shown in Figure \ref{fig1_SHG} (b) \cite{Hsu2014,Psilodimitrakopoulos2021}. Note that a twist angle of $\theta = 60^{\circ}$ results in completely destructive interference, fully suppressing the SHG response, while it is maximal for $\theta = 0^{\circ}$. Moreover, strong electric dipole SHG from centrosymmetric bilayer MoS$_2$ with $2-H$ stacking has been observed \cite{Shree2021, Zhang2023_shg_mos2_graphene} due to a resonance enhancement associated with inter- and intra-layer excitons, a phenomenon also expected to occur in heterobilayers hosting \moire{} excitons. Indeed, strong twist angle enhancement has been observed in MoS$_2$/WSe$_2$ devices at the intralayer hybridized exciton energies \cite{PhysRevB.105.115420}. This is highlighted in Figure \ref{fig1_SHG} (c). The SHG response is dominated by two main resonances at twice the energy of the A1s exciton of the MoSe$_2$ (1.92 eV) and WSe$_2$ (1.72 ev) monolayers. At those energies, the SHG polar plot at the heterostructure selectively reflects that of the corresponding monolayer. Comprehensive reviews of the enhancement of non-linear processes in twisted and heterostructures of TMDs can be found in Ref. \cite{Wen_nonlinear_2019,ZhouKrasnokHussainYangUllah_2022}. Thus, SHG has emerged as a key technique in the study of \moire{} heterostructures to confirm the desired twist angle quickly, versatilely, and non-disruptively. However, the extreme surface sensitivity of SHG requires clean and flat samples. Moreover, SHG is limited by the diffraction limit of light, making its deployment to small and complex stacks challenging.

Optical contrast of the \moire{} heterobilayers beyond subwavelength resolution ($\sim$ 20 nm) can be achieved by scanning near-field optical microscopy (SNOM) \cite{Vincent2021}. In a scanning SNOM (s-SNOM), the metallic AFM tip locally confines the electric light field. Thus, the excitonic spectrum of TMD heterostructures could be resolved by measuring the scattered light amplitude and phase as a function of position and energy \cite{PhysRevB.90.085136}. s-SNOM has been recently deployed to monolayers of MoSe$_2$ and WSe$_2$ and MoSe$_2$/WSe$_2$ heterobilayers to extract the excitonic resonance energy and interlayer hybridization with nanometer resolution \cite{Zhang2022} [Figure \ref{fig1_SHG} (d)-(e)]. However, s-SNOM is not free of sample-tip interactions that can modify the properties of the heterostructure, which can be deployed to control the electronic structure of \moire{} excitons. For example, the AFM tip from a SNOM has been used to introduce a vertical deformation in heterostructures of WSe$_2$/WS$_2$ at cryogenic temperatures and concomitantly measure a shift on the \moire{} intralayer exciton as a function of uniaxial pressure \cite{SNOM_pressure}, as shown in Figure \ref{fig1_SHG} (f). The exciton energy reduces by more than $\approx 40$~meV for vertical displacements that correspond to $\approx1.4$~GPa.

Optical characterization techniques with real space imaging capabilities are thus fundamental to studying the properties of \moire{} heterostructure and excitons. Moreover, their large versatility enables a multi-messenger approach combining different wavelengths and scanning techniques to characterize the excitation spectrum over a broad range of energies \cite{Sternbach2023_Multimessenger}.

\subsection{Real and reciprocal space imaging of \moire{} excitons using photoemission microscopes}\label{subsection:ARPES}

\begin{figure*}[t!]
\includegraphics[width=0.9\textwidth]{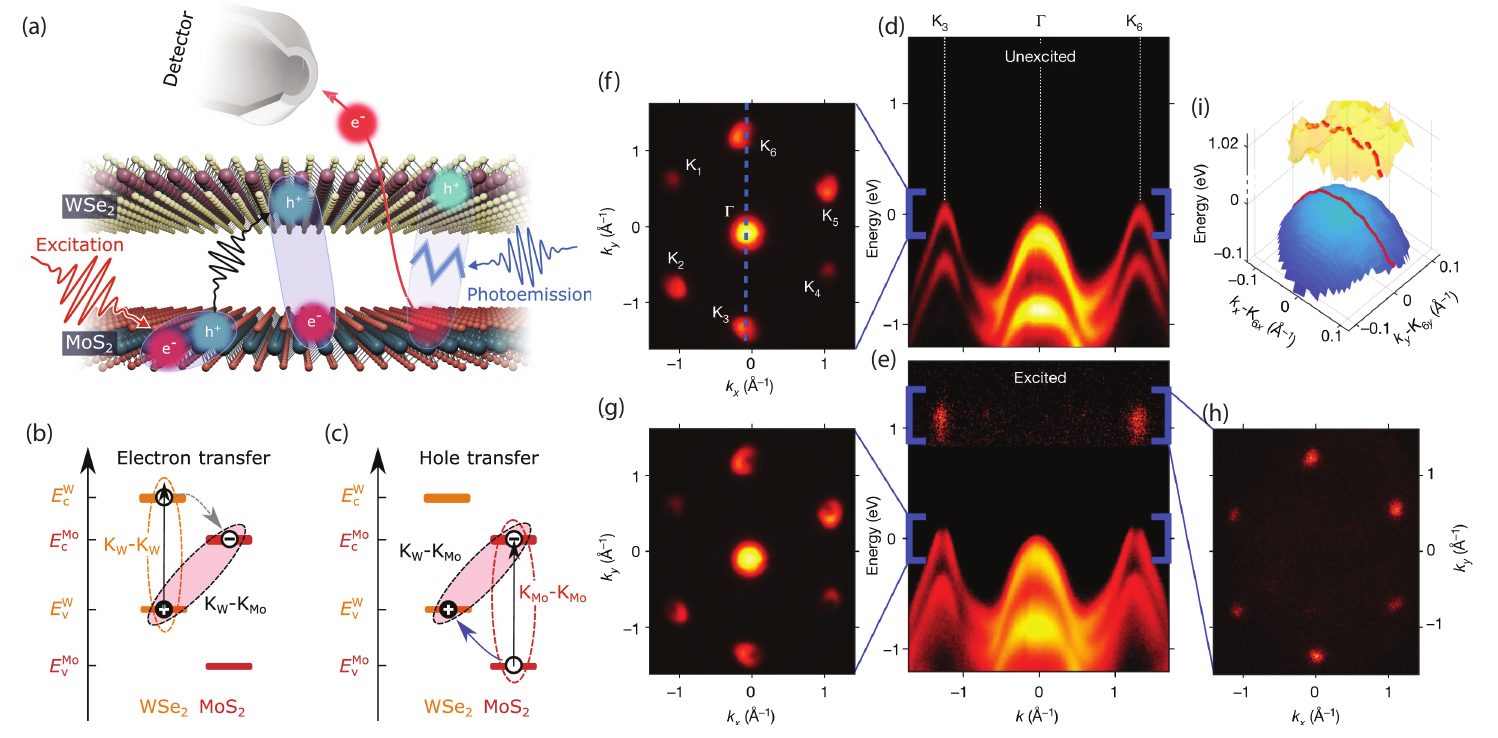}
\caption{(a) Schematic on the ARPES process in the exemplary WSe$_2$-MoS$_2$ heterostructure after optical excitation of excitons. A second extreme ultraviolet laser pulse breaks the exciton, and the ejected electrons are detected while the holes remain in the material. (b)-(c) Electronic band alignment in the heterostructure with the formation of interlayer excitons via electron or hole transfer, depending on the excitation condition. Energy–momentum cuts of the ARPES signal along the K–$\Gamma$ direction in the Brillouin zone (d) before and (e) after optical excitation of excitons in the WSe$_2$-MoS$_2$ heterostructure. The insets show the zoomed-in momentum-space images around the valence band minimum (f,g) and around the interlayer exciton energy (h). Note that the interlayer exciton signal has been enhanced by a factor of 50. (i) The ARPES signal from the ejected electrons exhibits the negative dispersion of the remaining holes. Figures adapted with permission from Ref. \cite{Bange2024,Karni2022}. 
\label{fig1_arpes}}
\end{figure*}

Angle-resolved photoemission spectroscopy (ARPES) can directly access a material-specific electronic band structure \cite{RevModPhys.93.025006}. In an ARPES measurement, an extreme ultraviolet probe pulse is applied to eject electrons from the material [Figure \ref{fig1_arpes} (a)]. Measuring the kinetic energy and the emission angle of the ejected electrons allows us to extract the energy and momentum of the initially excited electrons in the material \cite{Dani2023_Review}. The successful development of high-repetition-rate extreme ultraviolet light sources pawed the way for time-resolved ARPES measurements (tr-ARPES) on TMD materials. In a tr-ARPES experiment, a second pump pulse, typically in the visible or near-infrared range, enables sensitivity to the transient dynamics of photoexcited carriers with momentum resolution. For excitonic materials, such as TMDs, tr-ARPES measurements provide access to exciton binding energies and their momentum in the reciprocal space. This powerful technique has been demonstrated to directly visualize momentum-dark excitons in TMD monolayers \cite{Madeo2020} and the momentum-resolved charge transfer dynamics in twisted TMD heterostructures \cite{Schmitt2022,Bange2024}. While the first ARPES studies were performed on exciton dynamics in bulk TMD materials \cite{PhysRevLett.117.277201,PhysRevB.102.125417,Nat_Sci_Dong,cabo_2015}, more recent results have focused on the exciton physics of TMD monolayers enabled by the development of time-of-flight momentum microscopes \cite{Madeo2020,Wallauer_2021,PhysRevLett.130.046202}. For a more in-depth discussion of the techniques' details, we refer the reader to other recent reviews \cite{Dani2023_Review,RevModPhys.93.025006}.

We start by discussing how tr-ARPES can reveal the dynamics of excitons in TMDs with momentum resolution. The pump (excitation) pulse is tuned to create excitons in the material. The second pulse breaks up the Coulomb-bound electron-hole pairs, photo-ejecting the electrons while leaving a hole density behind [Figure \ref{fig1_arpes} (a)]. The ARPES signal of the photoemitted electrons reflects the excitonic band structure since the Coulomb-bound electrons and holes act as a strongly correlated system \cite{Menghini_2023}. In energy, the ARPES response is spectrally located one exciton binding energy below the conduction band. The excitonic ARPES signal appears at the momentum point corresponding to the ejected electron's reciprocal-space valley, and its dispersion is characterized by the negative curvature of the valence band of the remaining hole \cite{PhysRevB.100.205401,Man2021,Menghini_2023}.

The rest of this section is organized as follows. We first review the exciton dynamics in TMD homo- and heterobilayers. In the latter case, type-II band alignment favors the tunneling of an electron or hole into the opposite layer. This leads to spatially separated interlayer excitons \cite{Yong2019}, where the Coulomb-bound electrons and holes are located in different layers [Figure \ref{fig1_arpes} (b)-(c)]. Second, we introduce ultrafast dark field momentum microscopy a new technique for simultaneous nano-imaging and nano-spectroscopy of materials \cite{schmitt2023ultrafast} with a temporal resolution of 55 fs and a spatial resolution of $\approx 100$~nm. Finally, we discuss how the deployment of ultrafast dark field momentum microscopy has revealed local inhomogeneities of the dark excitons. 

\begin{figure*}[t!]
\includegraphics[width=\textwidth]{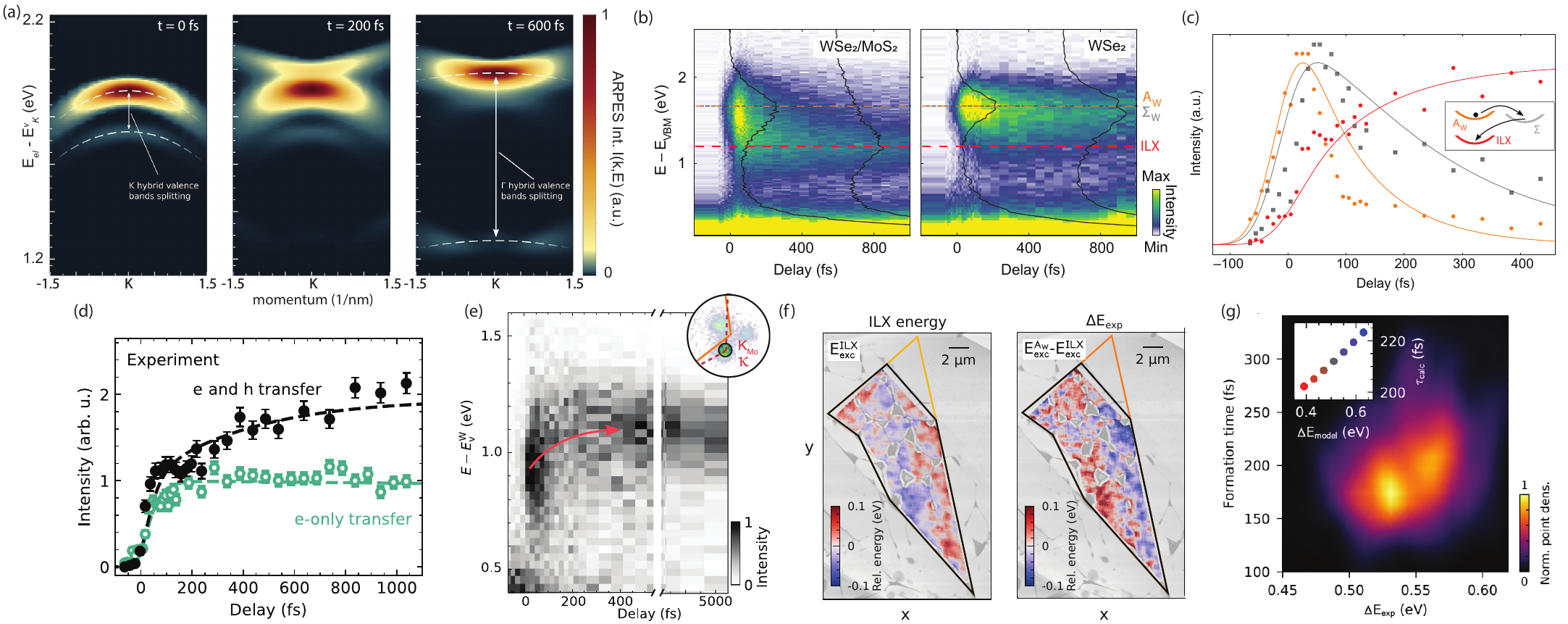}
\caption{ (a) Theoretical prediction for the hybrid exciton dynamics in the MoS$_2$ homobilayer showing the excitation of the nearly purely intralayer K-K excitons (0 fs), the formation of strongly hybridized $\Gamma_{\text{hyb}}$-K excitons (200 fs), and the thermalized hybrid exciton distribution (600 fs), respectively. The dashed lines denote the shifted split valence bands of the hybrid hole at the K and the $\Gamma$ point. (b) Momentum-integrated energy distribution for the heterostructure (left) and the WSe$_2$ monolayer (right). (c) Charge-transfer dynamics from optical excitation of excitons in the WSe$_2$ monolayer (K$_{\text{W}}$-K$_{\text{W}}$, orange line) via phonon-driven scattering into the hybrid dark excitons (K$_{\text{W}}$-$\Lambda_{\text{hyb}}$, grey line) to the energetically lowest interlayer exciton states (K$_{\text{W}}-$K$_{\text{Mo}}$, red line). The points represent experimental data, and solid lines are the theoretical predictions. (d) Direct comparison of the charge transfer dynamics when the heterostructure is excited resonant to the excitons either in the WSe$_2$ layer (green circles, 1.7 eV) or in the MoS$_2$ layer (black circles, 1.9 eV). While the first excitation scenario leads to an electron transfer only, the latter gives rise to a combined electron- and hole-transfer. (e) Time evolution of the energy distribution filtered at the momentum region of the K valley in MoS$_2$ layer for the excitation at 1.9 eV. As the interlayer exciton is formed, the photoelectron energy shows an unintuitive blue shift. (f) Spatial heatmaps illustrate the variation of the interlayer exciton (ILX) energy and the energy difference $\Delta E_{\text{exp}}$ to the optically excited interlayer exciton, respectively. (g) The interlayer exciton formation time is plotted as a function of the energy difference $\Delta E_{\text{exp}}$. The inset shows a theoretical prediction confirming qualitatively the measured correlation between the formation time and the energy difference. Figures adapted with permission from Ref. \cite{Schmitt2022,schmitt2023ultrafast,Menghini_2023,Bange2024}
\label{fig2_arpes}}
\end{figure*}

In a twisted WSe$_2$-MoS$_2$ heterostructure, ARPES signatures of the bound electron and hole pairs in an interlayer exciton have been observed [Figure \ref{fig1_arpes} (d)-(i)]. This was enabled by the high signal-to-noise ratio and high-quality samples with narrow initial valence band line widths \cite{Karni2022}. The unexcited band structure of the twisted WSe$_2$-MoS$_2$ heterostructure is characterized by two spin-split valence band maxima of the WSe$_2$ layer at the K valley and a single broad valence band of the MoS$_2$ layer  [Figure \ref{fig1_arpes} (d) and (f)]. After optically exciting the material with photons resonant at the WSe$_2$ intralayer exciton, clear signatures of interlayer excitons about 1 eV above the valence band maximum are found [Figure \ref{fig1_arpes} (e)]. Concomitantly, the signal around the WSe$_2$ valence band maximum becomes slightly depleted at the K and K’ points [Figure \ref{fig1_arpes} (g) and (h)]. This was assigned to the creation of holes in WSe$_2$ during photoexcitation. Finally, the interlayer exciton signature at 1 eV exhibits a negative dispersion resembling the negative curvature of the WSe$_2$ valence band [Figure \ref{fig1_arpes}(i)]. This anomalous dispersion has been theoretically predicted \cite{PhysRevB.100.205401} and is a hallmark of the excitonic origin of the observed ARPES signal. 

Recently, it was shown that tr-ARPES is an ideal technique to directly visualize hybrid exciton states in \moire{} heterostructures \cite{Menghini_2023}. Hybrid \moire{} excitons refer to excitonic states formed by a mixture of intra- and interlayer exciton states due to the direct tunneling between two TMD layers. In other words, electrons and holes are in superposition in both layers of the \moire{} heterostructures. The formation of such hybrid excitonic states is facilitated in naturally stacked TMD homobilayers, making them the dominating excitons of interlayer character as first observed in WSe$_2$/WS$_2$ \moire{} heterostructures \cite{Naik2022}.

Thus, we first review a tr-ARPES study in MoS$_2$ homobilayers \cite{Menghini_2023}. Figure \ref{fig2_arpes} (a) illustrates how the momentum- and energy-resolved ARPES signal evolves in time. During the optical excitation, the spectrum is predicted to be characterized by a pronounced single peak stemming from the optically excited intralayer exciton [Left panel of Figure \ref{fig2_arpes} (a)]. On a sub-100 fs timescale, these excitons scatter via emission of phonons to the energetically lower hybridized $\Gamma_{\text{hyb}}$-K excitons and the ARPES signal shows signatures of both K-K intralayer and hybridized $\Gamma_{\text{hyb}}$-K excitons [Middle panel of Figure \ref{fig2_arpes} (a)]. After $\approx 600$~fs, an equilibrium distribution is reached with $\Gamma_{\text{hyb}}$-K excitons carrying nearly the entire population [Right panel of Figure \ref{fig2_arpes} (a)]. The ARPES signal is predicted to display two peaks, one slightly above the K-K exciton and one red-shifted by more than 600 meV. This work revealed that the energetically lowest states are the hybridized momentum-dark $\Gamma_{\text{hyb}}$-K excitons \cite{Menghini_2023}. Here, the hole at the $\Gamma$ valley is strongly delocalized between the layers. During the ARPES process, the hybrid exciton breaks up into an ejected free electron from the K valley and a superposition of hybrid holes. This gives rise to a double-peak signal reflecting the superposition of the remaining holes between the two hybrid valence bands at the $\Gamma$ point.

We proceed to review hybrid excitons in \moire{} heterobilayers. In the following, the first/second capital letter denotes the position of the hole/electron in the reciprocal space (K, $\Lambda$ valley), while the subindex stands for the layer (W for WS$_2$ and Mo for MoS$_2$). The left panel of Figure \ref{fig2_arpes} (b) shows the momentum-integrated energy distribution of the tr-ARPES signal in WSe$_2$-MoS$_2$ \moire{} heterostructures \cite{Schmitt2022}. The highest contribution at early delay times occurs at around 1.7 eV (orange dashed line), corresponding to the energy of the optically excited K$_{\text{W}}-$K$_{\text{W}}$ intralayer excitons in the WSe$_2$ layer. A transition to a second long-lived peak at the lower energy of about 1.1 eV (red dashed line) occurring on a femtosecond timescale is observed. This peak has been identified as the photoemitted electronic contribution of the K$_{\text{W}}$-K$_{\text{Mo}}$ interlayer hybrid exciton (hole at the K valley of the WSe$_2$ layer and electron in the K valley of the MoS$_2$) [Figure \ref{fig1_arpes}(b)]. The same measurement was performed on a WSe$_2$ monolayer to support this assignment, as shown in the right panel of Figure \ref{fig2_arpes} (b). No spectral weight is observed in the momentum and energy space region associated with the interlayer excitons in the monolayer, highlighting the hybrid nature of the exciton in WSe$_2$-MoS$_2$ heterostructures.

The 1D plots in Figure \ref{fig2_arpes} (c) summarize the observed charge transfer dynamics shown in the 2D plots of Figure \ref{fig2_arpes} (b). The optically excited intralayer K$_{\text{W}}-$K$_{\text{W}}$ excitons in the WSe$_2$ layer cannot efficiently scatter directly to the energetically lowest K$_{\text{W}}-$K$_{\text{Mo}}$ interlayer exciton states. The K$_{\text{W}}-$K$_{\text{W}}$ excitons rather scatter first with phonons to the momentum-dark K-$\Lambda$ excitons, where the electrons at the $\Lambda$ valley are delocalized over both layers. In a second step, these hybridized excitons scatter into the K$_{\text{W}}-$K$_{\text{Mo}}$ interlayer excitons, where the electron has been transferred to the MoSe$_2$ layer \cite{Meneghini_nat_sci}. We note that the direct scattering process is prevented by the weak wave function overlap at the K valley. In contrast, the electronic wave function at the $\Lambda$ valley has large contributions at the outer selenium atoms, resulting in much more efficient hybridization of K-$\Lambda$ states \cite{Merkl2020}.
 
Tr-ARPES has also revealed the hole transfer process mechanism across the WSe$_2$-MoS$_2$ interface \cite{Menghini_2023}. The K$_{\text{Mo}}$-K$_{\text{Mo}}$ intralayer excitons in the MoS$_2$ layer can be resonantly excited at 1.9 eV [Figure \ref{fig2_arpes} (d)]. However, this still leads to a partial electron transfer due to the off-resonant excitation of the energetically lower K$_{\text{W}}$-K$_{\text{W}}$ excitons (1.7 eV) in the WSe$_2$ layer. To distinguish the impact of the electron and hole transfer, the temporally resolved formation of K$_{\text{W}}$-K$_{\text{Mo}}$ interlayer excitons have been studied for both scenarios, i.e. excitation at 1.7 eV (only electron transfer expected) and excitation at 1.9 eV (electron and hole transfer expected). For the electron-only transfer process, the interlayer excitons are formed quickly ($\sim$ 35~fs), and the ARPES signal saturates on the sub-200 fs timescale [Figure \ref{fig2_arpes} (d)]. In contrast, the joint electron- and hole-transfer is found to be much slower, with a formation time of approximately 3~ps. This can be traced back to the different relaxation pathways of optically excited K$_{\text{Mo}}$-K$_{\text{Mo}}$ and K$_{\text{Mo}}$-K$_{\text{Mo}}$ excitons. The degree of the hybridization of the intermediate hybrid states and the energetic offset to the initial and final states determine the efficiency of the charge transfer process. While the first scatter via hybridized $\Gamma_{\text{hyb}}$-K$_{\text{Mo}}$ excitons, the latter scatter via hybridized K$_{\text{W}}$-$\Lambda_{\text{hyb}}$ states towards the energetically lowest K$_{\text{W}}$-K$_{\text{Mo}}$ interlayer excitons. In principle, no difference in the ARPES response is expected whether the measured photoelectrons stem from the break-up of the intralayer K$_{\text{Mo}}$-K$_{\text{Mo}}$ or interlayer K$_{\text{W}}$-K$_{\text{Mo}}$ excitons as in both cases the electron remains in the MoS$_2$ layer while the hole transfers across the interface [Figure \ref{fig2_arpes} (d)]. On the other hand, a blue shift of about 170 meV of the ARPES signal is found at increasing delay times [Figure \ref{fig2_arpes} (e)] instead of the redshift expected in optically excited intralayer excitons due to scattering into the energetically lower interlayer states. Auger processes and energy renormalization could be excluded as possible explanations for the unexpected blue shift. Thus, the shift towards higher energies can only be understood as a hallmark of the correlated nature of the hybridized electron-hole pairs. 

Finally, we review ultrafast dark field momentum microscopy, introduced as a new technique for simultaneous nano-imaging and nano-spectroscopy of materials \cite{schmitt2023ultrafast}. Typically, tr-ARPES experiments do not spatially resolve the carrier dynamics, but they average over large sample areas of typically 10$\mu$m or more. Thus, they cannot access spatial inhomogeneities, dielectric disorder, strain gradients, or local lattice reconstructions. Ultrafast dark field momentum microscopy has been applied to the study of spatial heterogeneity in a WSe$_2$-MoS$_2$ heterostructure and its impact on the energy and the formation time of hybrid interlayer excitons. Distinct spatial heterogeneity is found even in optically flat areas of the sample exhibiting high-quality photoemission spectra [Figure \ref{fig2_arpes} (f)]. Positioning the dark-field aperture on the momenta of the interlayer excitons provides access to the spatially and spectrally resolved evolution of the interlayer exciton energy. The nanoscale variation of the latter, as well as its difference $\Delta E_{\text{exp}}$ to the energy of the optically excited intralayer exciton, is shown in spatial heat maps in Figure \ref{fig2_arpes} (f). A correlation was observed between the interlayer exciton formation time $\tau$ and the nanoscale variation of the exciton energy $\Delta E_{\text{exp}}$. $\tau$ gets larger with the increasing $\Delta E_{\text{exp}}$ [Figure \ref{fig2_arpes} (g)]. This observation is qualitatively confirmed by the many-particle theory (inset), which could trace this correlation back to nanoscale variations of the strength of interlayer hybridization in the WSe$_2$-MoS$_2$ bilayers. This study revealed how spatial inhomogeneities directly impact the formation timescale of interlayer excitons by influencing the efficiency of the phonon-driven charge transfer to the intermediate hybridized states.

Despite requiring careful sample preparation, including the need for vacuum annealing to remove surface adsorbates and for electrical isolation from the substrate while avoiding sample charging, tr-ARPES and ultrafast momentum microscopy have emerged as unique tools for understanding the formation of \moire{} excitons and the band structure of \moire{} heterostructures.

\subsection{Atomic relaxation and its influence on \moire{} excitons}\label{subsection:STM}
\begin{figure}[ht!]
\includegraphics[width=\columnwidth]{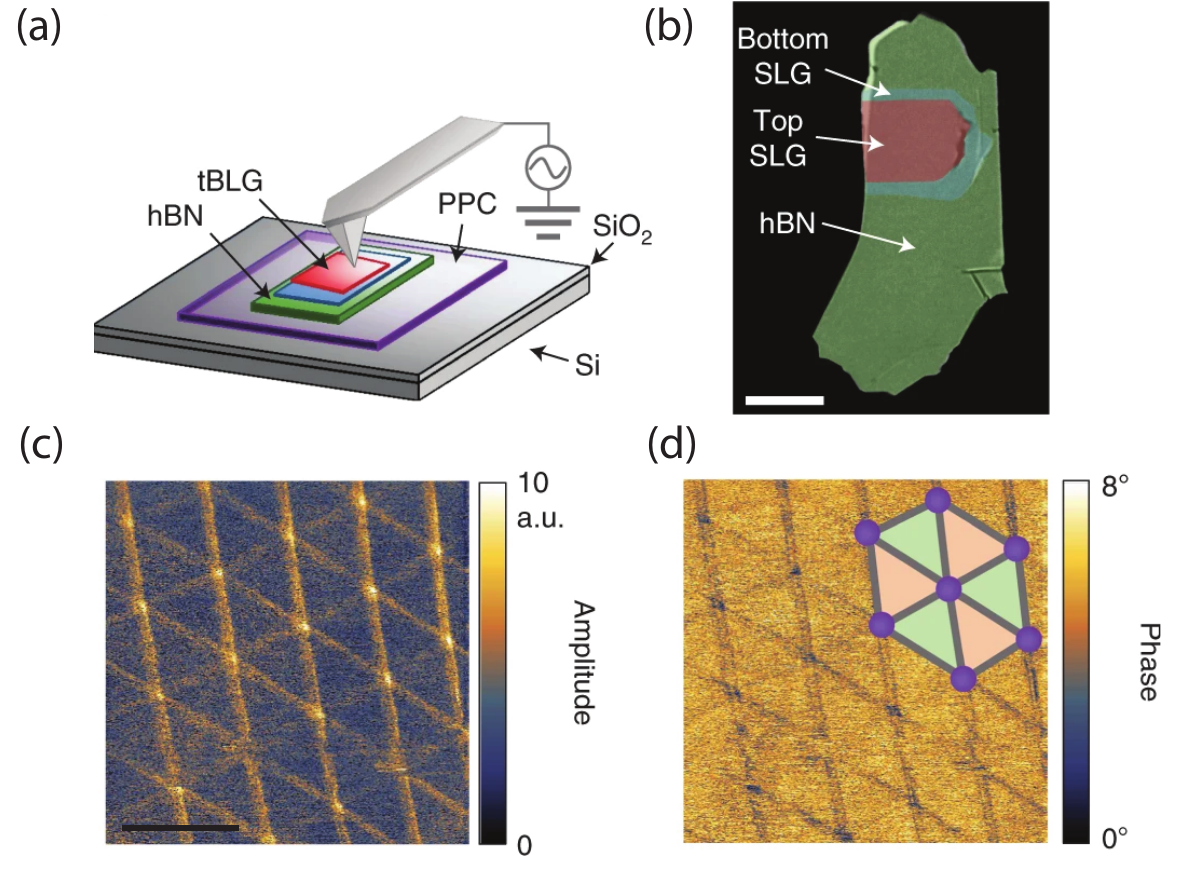}
\caption{(a) Schematic of a piezoresponse force microscopy (PFM) experiment on twisted bilayer graphene (tBG) prepared using standard polymer stamp dry-transfer technique. (b) Optical micrograph of the tBLG sample. Each of the single graphene layers (SLG) are highlighted. The scale bar corresponds to $10 \mu$m. (c) PFM amplitude and (d) phase revealing the \moire{} pattern superlattice. The overlay in (d) guides the eye to identify the stacking AA sites in purple,  domain walls in grey and AB/BA domains in green and orange. The scale bar is 500~nm. Adapted with permission from \cite{McGilly2020}}
\label{fig_PFM}
\end{figure}

\begin{figure*}[t!]
\includegraphics[width=\textwidth]{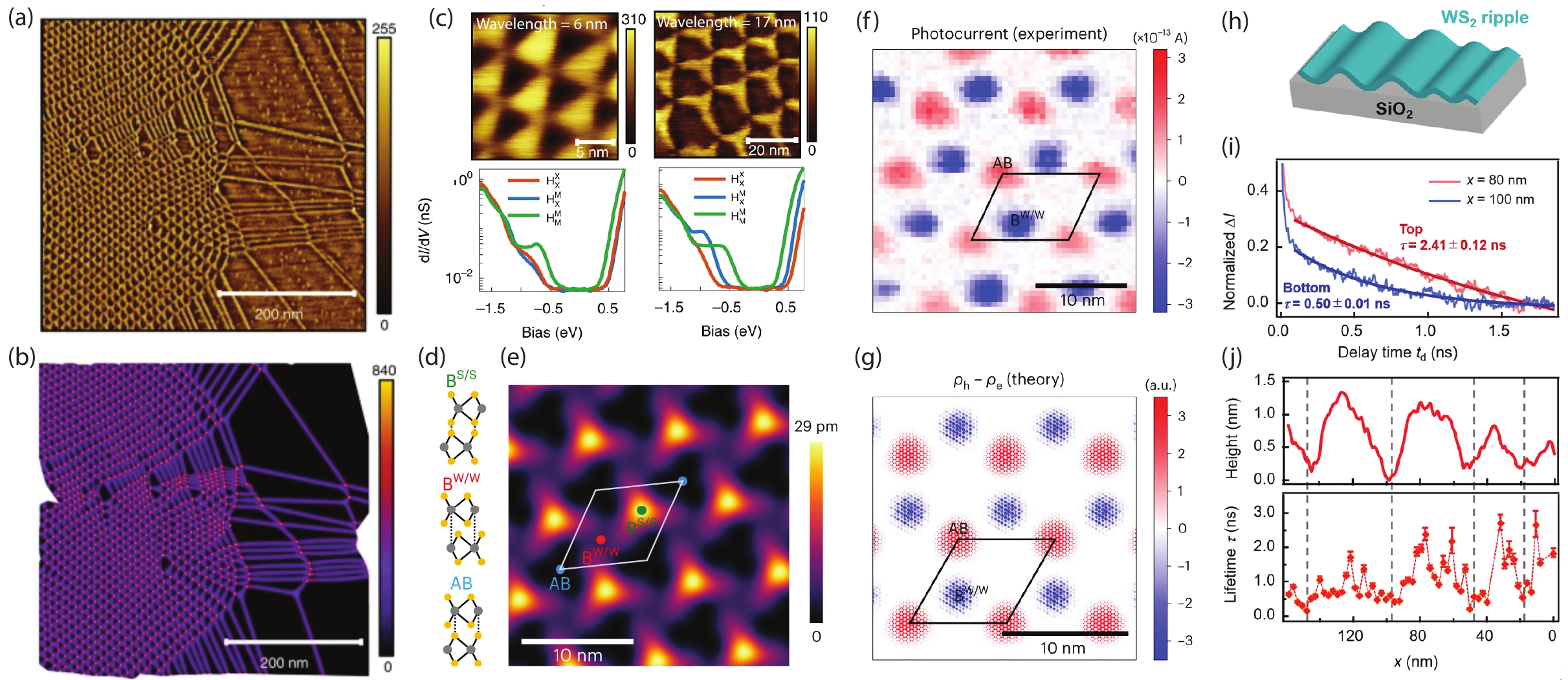}
\caption{ (a) Large STM topography in a $1.7^{\circ}$ WSe$_2$/MoSe$_2$ heterostructure showing a spatially dependent \moire{} periodicity. (b) Calculated stacking energies for the relaxed atomic energy reflecting the relation between the \moire{} pattern and strain. (c) (Top) STM topography for two \moire{} patterns with 6~nm (left) and 17~mm (right) wavelength. (Bottom) $dI/dV$ curves at three H stacking configurations ($H^{M}_{M}$, $H^{M}_{X}$ and $H^{X}_{X}$). The band gap changes with the \moire{} pattern wavelength. (d) Illustration of the possible local stacking structures in a near-58$^{\circ}$ twisted WS$_2$ bilayer sample. (e) Calculated topographic variation twisted WS$_2$ bilayers at $B^{W/W}$ and AB stacking sites. (f) Comparison of the experimentally observed spatial dependence of the charge transfer excitonic photocurrent sign and (g) GW-Bethe–Salpeter equation calculations. (h) Schematic of strain-induced ripples in WS$_2$ on SiO$_2$. (i) Intralayer exciton lifetime dependence at the top and bottom of a strain-induced ripple in WS$_2$. (j) Dependence of the exciton lifetime across a ripple with multiple crests. The exciton lifetime increases at the peak of the ripple. Adapted with permission from \cite{Shabani2021,Li2024,Mogi2022}.}
\label{fig1_local}
\end{figure*}

In recent efforts to build new 2D heterostructures combining distinct electronic and magnetic ground states from different v.d.W. materials \cite{geim2013van}, two fundamental real space properties of these compounds have been predominantly ignored: (i) their lattice constants are extremely dependent on local changes of the composition and (ii) the atomistic positions in v.d.W. materials are not rigid but rather follow a membrane-like behavior. The second point becomes most prominent in large \moire{} unit cells formed by small twist angles between adjacent layers or by small lattice constant mismatches in heterostructures. In this scenario, the relaxation of atomistic positions in and out of the plane can minimize the energy of the system. Thus, the reconstructed atomistic registry might be very different from that expected for unrelaxed, rigid heterobilayers with severe consequences for electronic, excitonic, and mechanical properties \cite{PhysRevB.98.224102,Halbertal2022,Susarla_hyperspectral}. It has recently been realized that atomic corrugation plays a crucial role in the flat bands and energy gaps of twisted bilayer graphene at the magic angle \cite{Uchida_PhysRevB.90.155451,Lucignano_PhysRevB.99.195419}. Moreover, vibrational stacking re-arrangements, known as \moire{} phonons, have been highlighted as an unconventional way to boost exciton diffusion \cite{rossi2023phasonmediated,Qian_2024_small}. Thus, it is fundamental to characterize experimentally the local stacking arrangements.

Piezoresponse force microscopy (PFM) is an extension of atomic force microscopy (AFM) that has emerged as a universal and easily implemented technique to map the real space variation of the \moire{} potential \cite{McGilly2020}. PFM measures the amplitude and phase of the electromechanical response of a material to local deformations induced by an alternating current (AC) bias between its surface and the AFM tip [Figure \ref{fig_PFM} (a)]. The resolution figure of merit in a PFM experiment is given by the contact radius of the AFM, which is generally around $\approx 5$~nm. PFM is a highly versatile tool that can quickly identify the local variations of the \moire{} potential. When deployed to \moire{} superlattices, irrespectively of the inversion symmetry of the constituent mono-layers, PFM can track the changes in the flexoelectric response due to inherent strain gradients in the heterostructure leading to polar domain walls in a non-polar background. \cite{McGilly2020}. This is exemplified in Figure \ref{fig_PFM} (c) and (d), where the PFM amplitude and phase change show the domain wall array as a result of the \moire{} pattern in a twisted bilayer graphene (tBLG) sample [Figure \ref{fig_PFM} (b)]. Despite the centrosymmetric nature of tBLG, a large PFM response is found at the AA stacking site and domain walls, while AB domains give a negligible response. 

\begin{figure}[t!]
\includegraphics[width=\columnwidth]{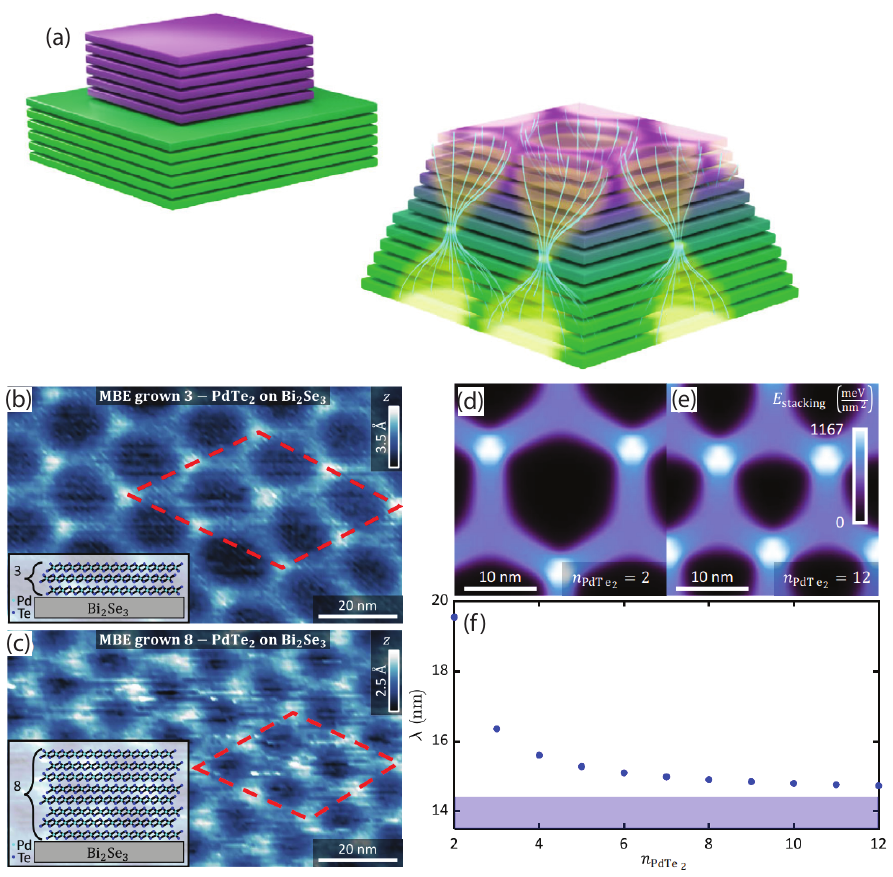}
\caption{ (a) Schematic comparison of the lego-like and continuous layer-dependent interpretation of 2D heterostructures. (b)-(c) STM topography of 3 and 8 ML, respectively, of PdTe$_2$ on Bi$_2$Se$_3$ heterostructures. (d)-(e) Calculated \moire{} pattern for eight, (d) and twelve, (e), PdTe$_2$ layers after the atomic positions have been relaxed. (f) Calculated \moire{} potential wavelength dependence on the number of PdTe$_2$ layers. Adapted with permission from \cite{PhysRevX.13.011026}}
\label{fig2_local}
\end{figure}

Going beyond nm resolution, scanning tunneling microscopy and spectroscopy experiments (STM/S) emerge as the only techniques with the required atomic and energy resolution to observe changes to the excitonic landscape due to atomic relaxation. For an in-depth review of the STM/STS technique and its application to quantum materials, we refer the reader to previous reviews \cite{Bian2021,Yin2021}. Additionally, early STM/S measurement on \moire{} heterostructures \cite{Zhang2017, Zhang2020,Li2021} were reviewed in Ref. \cite{Huang_Excitons_moire_superlattices}.

We now discuss examples of the effects of local strain in the electronic structure of \moire{} heterobilayers, as revealed by STM/S. In small-angle heterostructures of WSe$_2$/MoSe$_2$ STM experiments revealed spatially varied \moire{} period up ranging from $5$ to $17$~nm wavelengths over a $500 \times 500 $ nm$^2$ \cite{Shabani2021} [Figure \ref{fig1_local} (a)]. The resulting strain associated with the \moire{} structure induces systematic large variations on the \moire{} potential depths that can be recovered by calculations considering the effects of generalized forces acting on a stack of membranes \cite{Carr2018} [Figure \ref{fig1_local} (b)]. Moreover, STS measurements reveal local changes to the near $E_F$ states with the \moire{} potential wavelength, with the electronic gap narrowing at longer wavelengths [Figure \ref{fig1_local} (c)]. Additionally, in $3^{\circ}$ twisted bilayers of WS$_2$ STS revealed the existence of flat bands associated with localized states that cannot be accounted for with first-principle calculations but that can be recovered when the atomic displacements due to strain are included \cite{Molino_NanoLetters}. STM has also been deployed to observe changes of more than $10\%$ in the \moire{} period of heterostructures of molecular-beam epitaxy (MBE) grown multilayer PdTe${_2}$ on Bi$_2$Se$_3$ [Figure \ref{fig2_local} (b)] \cite{PhysRevX.13.011026}.   

In studying intra- and interlayer \moire{} excitons, it is also vital to understand and characterize the precise atomistic registry as it will modify the specific interactions and relaxation pathways. Laser-STM experiments in which STM/STS measurements are performed while a continuous-wave laser shines on the sample, enabling resolving the electronic structure of \moire{} excitons with sub-nm resolution. Laser-STM experiments in twisted bilayer WS$_2$ revealed nanoscale modulation in the \moire{} excitons \cite{Li2024}. The STM topography maps in these heterostructures are characterized by a \moire{} pattern with a 9~nm period [Figure \ref{fig1_local} (e)]. STS measurements show clear differences in the electronic band structure at B$^{W/W}$, direct stacking of W atoms of different layers, and AB stacking sites reflecting the deep \moire{} potential in twisted WS$_2$ [Figure \ref{fig1_local} (d)]. Upon turning the laser on, the emerging excitonic photocurrent displays different behavior for both sites. When the bias voltage between tip and sample is set in the region $-2 V < V_{bias} < 1 V$, the photocurrent is concomitantly positive in the B$^{W/W}$ sites and a negative in the AB stacking sites [Figure \ref{fig1_local} (f)], indicating a dominant electron and hole contribution respectively. This phenomenology was explained by GW-Bethe–Salpeter equation calculations, including the \moire{} potential and electron-hole Coulomb interactions, which indicate the formation of in-plane charge transfer (ICT) \moire{} excitons. Thus, the STM tip-position photocurrent can be explained as a result of the lateral separation of electrons and holes in an ICT exciton [Figure \ref{fig1_local} (g)] \cite{Li2024}.

Optical pump-probe STM setups with picosecond resolution could enable studying the atomic dependence of exciton dynamics \cite{tr-STM_review}. The challenge in tr-STM measurements is to remove thermal fluctuations associated with the optical pump-induced expansion of the atomic tip and sample, which can result in large artifacts in the tunneling currents. The development of shaken-pulse-pair-excited STM, in which the pump beam is electronically modulated by an external trigger acting as the reference signal for lock-in amplifier detection \cite{Mogi_2019,Takeuchi_2002}, has enabled the required high-signal-to-noise ratios and time resolution to study non-equilibrium dynamics with atomic resolution \cite{Iwaya2023}. Recently, tr-STM has been deployed to study the spatial dependence exciton dynamics in $7.6^{\circ}$ heterobilayers of WS$_2$/WSe$_2$ \cite{Mogi2022} near grain boundaries and strain-induced ripples [Figure \ref{fig1_local} (h)]. The authors revealed an almost fivefold enhancement of the exciton lifetime between the top and bottom of the ripple, suggesting that interaction with the substrate phonon modes can lead to shorter excitonic lifetimes [Figure \ref{fig1_local} (i) - (j)]. Similarly, substrate-induced strain near grain boundaries also leads to three times faster excitonic relaxation than at the grain boundary \cite{Mogi2022}. However, the effect diminishes rapidly in space, $\approx 8$~nm away from the grain boundary, where long exciton lifetimes are recovered \cite{Mogi2022}.

Theoretical modeling of the atomic relaxation requires properly considering the multi-layered nature of v.d.W. heterostructures. It has been shown that the strain fields emerging from \moire{} defects can penetrate deep into a v.d.W. material stack~\cite{cook2022moire,PhysRevX.13.011026}. This departs from the Lego-like interpretation of v.d.W. heterostructures towards a continuous layer-dependent interpretation of defects and strain [Figure \ref{fig2_local} (a)]. However,  a theoretical description of v.d.W. heterostructures with large \moire{} periods and/or many layers on the atomic scale is computationally out of reach. As a consequence, sophisticated coarse-grained models were developed that address atomic relaxation at large length scales. When the \moire{} lattice spans many unit cells of a single layer, the local atomic stacking registry and in- and out-of-plane strain fields can be modeled by considering each single layer as a membrane ~\cite{Carr2020,Carr2018,PhysRevX.13.011026,PhysRevLett.125.116404,Cazeaux2020,Halbertal2021}. In such membrane models, each layer of the v.d.W. stack is allowed to relax individually. The total energy functional is calculated as a sum of the generalized stacking fault energy, which accounts for lattice mismatch at stack interfaces, and the elastic energy of each membrane, which accounts for the in- and out-of-plane strain fields. The parameters for both the elastic and the stacking fault energy are readily obtained from \emph{ab-initio} simulations of the individual constituent layers and interfaces~\cite{PhysRevX.13.011026}. The multi-layered relaxation process in v.d.W. heterostructures can heavily affect electronic and, thus, excitonic properties. Therefore, the theoretical modeling of relaxation, e.g., by the aforementioned membrane models, is an essential ingredient in describing these properties in moir{\'e} structures even far away from the interface~\cite{PhysRevX.13.011026} or in novel interfaces between two three dimensional crystals ~\cite{Mullan2023} .

These experimental and theoretical studies exemplify that it is essential to understand the atomic relaxation mechanisms of \moire{} heterostructures to fully understand their emergent excitonic properties. In a second step, one might consider using long-ranged relaxation even for engineering-by-stacking purposes.

\section{Outlook}

New technological applications are emerging as our understanding of \moire{} physics progresses. Semiconductor \moire{} superlattices and Kekul\'e/\moire{} superlattices offer new prospects for practical quantum devices \cite{Ye2023_Kekule}. For an extensive review of the applications of \moire{} optoelectronics, we refer the reader to Ref. \cite{Du_Science_2023} and Ref. \cite{Ciarrocchi_Excitonic_devices_vdWheterostructures,Qi_review_2022}. In the following, we highlight a few functional implementations in an early stage of development:

\textit{Optics, photonics, and optoelectronic devices:} \moire{} excitons have been shown to play a pivotal role in developing tunable optoelectronic devices across various heterostructures and materials \cite{Behura2021,pnas_interplay_defffect_moire,brem_2020,Liu2022}. Moir{\'e} excitons serve as a platform for optoelectronic modulation, offering efficient control over light absorption, emission, and transmission essential for devices like optical switches, photodetectors, and modulators, ensuring high-speed operation and low energy consumption \cite{Ramos_2022_PL_enhancement,Yu2017}. For example, \moire{} excitons enhance device efficiency in WS$_2$/graphene heterostructures by promoting efficient carrier extraction and utilization with high sensitivity and signal-to-noise ratios \cite{Trovatello2022}. Integration of \moire{} excitons into photonic circuits enables the development of compact, scalable systems for telecommunications, sensing, and quantum information processing. Specifically, MoS$_2$/FePS$_3$ heterostructures have the potential for new integrated photonics \cite{Duan2022_Self-Driven_Broadband}, offering multifunctional devices with advanced performance and functionality \cite{Onga2020_AFM_semiconductor}. Novel excitonic devices exploiting strong light-matter interactions and long-range effects will result in novel excitonic transistors, photodetectors, LEDs, and solar cells with enhanced efficiency and tunability.

\textit{Quantum Information Processing:} The long coherence times and strong interactions of \moire{} excitons position \moire{} devices as promising candidates for quantum information processing \cite{Huang_Excitons_moire_superlattices}. Moreover, \moire{} excitons serve as valuable sources of quantum light, capable of generating non-classical light states like single photons and entangled photon pairs \cite{Baek2020,Yu2017}. Thus, these microscopic properties could be leveraged toward effectively processing quantum information in quantum computing, quantum communication, and quantum cryptography applications. 

\textit{Other devices and applications}: \moire{} excitons offer significant potential for enhancing the efficiency and functionality of various energy, sensing and detection devices \cite{Ciarrocchi_Excitonic_devices_vdWheterostructures,Du2023}. Their strong light-matter interaction in energy harvesting and photovoltaics allows for efficient photon absorption and higher photoconversion efficiencies \cite{Duan2022_Self-Driven_Broadband}. Additionally, their tunability enables optimization across different wavelengths, making them suitable for diverse photonic applications. In broadband photodetection, MoSe$_2$/FePS$_3$ heterostructures demonstrate a broad spectral response, enhancing versatility for applications like imaging and spectroscopy \cite{Onga2020_AFM_semiconductor}. These photodetectors operate self-driven due to efficient carrier generation and transport, resulting in low power consumption and enhanced portability. Moreover, \moire{} exciton-based photodetectors exhibit fast response times and low noise performance, crucial for high-speed data acquisition and sensitive detection of weak optical signals in noisy environments. Enhanced light-matter interaction in \moire{} excitons, particularly in Kekul{\'e}/\moire{} superlattices and 2D TMD materials, improves optical absorption and emission properties, benefiting applications like optical sensing and imaging \cite{Ye2023_Kekule}. Furthermore, \moire{} excitons facilitate efficient hot carrier transfer in WS$_2$/graphene heterostructures, enabling rapid carrier dynamics and high-speed device operation essential for photodetectors and communication systems. Additionally, \moire{} excitons show potential in magnetic field sensing applications, especially in antiferromagnet-semiconductor heterostructures \cite{Onga2020_AFM_semiconductor}. Their sensitivity to external magnetic fields enables high-resolution and sensitive detection, making them suitable for various magnetic imaging, navigation, and magnetometry applications. Finally, the engineering of strong interlayer excitonic correlation in MoSe$_2$/hBN/WSe$_2$ heterostructures with equal electron and hole densities can lead to condensates at low-temperatures \cite{qi2023perfect,nguyen2023perfect}. This state is characterized by perfect Coulomb drag, in which a charge current in one layer is balanced by an equal but opposite current in the other that could be deployed in dissipationless excitonic transport devices.

This review highlights current efforts to concomitantly deploy theoretical and experimental efforts to characterize the structural variation of the \moire{} potential and collective excitations. However, it is fundamental to keep furthering our understanding of the role of spatial inhomogeneities of the \moire{} potential and local atomic relaxations on the electronic properties of \moire{} heterostructures towards the future deployment of novel technology based on \moire{} devices. In particular, how unavoidable defects introduced during the fabrication of \moire{} heterostructures \cite{Rhodes2019} affect the properties of \moire{} excitons remains largely unexplored. For example, recent experimental results have identified the presence of defect traps able to localize the exciton in MoSe$_2$/WSe$_2$ heterobilayer \cite{Fang2023}. Additionally, atomic defects can enable new electron-phonon coupling pathways in WS$_2$/graphene heterostructures \cite{Liu2022visualizing}. First principle calculations have proposed defect engineering as a promising route towards modulating the excitonic properties of \moire{} heterostructures \cite{pnas_interplay_defffect_moire}. While defect engineering is still in its infancy, it opens an alternative approach towards engineering the properties of v.d.W. \moire{} heterostructures \cite{Rakib2022}. Above all, it is clear that we are currently approaching a very promising future in which 2D v.d.W. \moire{} physics \cite{Andrei2021} gives rise to real-world technological applications.

\section{Acknowledgements}
We thank  Lennart Klebl for his valuable input. D.M.K gratefully acknowledges support from the DFG through SPP 2244 (Project No. 443274199) and by the Max Planck-New York City Center for Nonequilibrium Quantum Phenomena. E.M. has received funding from the Deutsche Forschungsgemeinschaft (DFG) via CRC 1083. S.K. acknowledges support provided by the National Science Foundation through the ExpandQISE award No. 2329067 and the Massachusetts Technology Collaborative through award number 22032.

%\bibliography{biblio}
%merlin.mbs apsrmp4-1.bst 2010-07-25 4.21a (PWD, AO, DPC) hacked
%Control: key (0)
%Control: author (3) reversed first dotless
%Control: editor formatted (0) differently from author
%Control: production of article title (0) allowed
%Control: page (1) range
%Control: year (0) verbatim
%Control: production of eprint (0) enabled
%

\end{document}